\documentclass[aps,amsmath,amssymb,twocolumn, superscriptaddress,preprintnumbers]{revtex4}
\pdfoutput=1
\usepackage{latexsym}
\usepackage{graphicx}
\usepackage{amsthm}
\usepackage{float}
\usepackage{amsmath}
\usepackage{graphicx,subfigure}
\usepackage{hangcaption,fancybox,feynmp}
\usepackage{indentfirst}
\usepackage{epsfig}
\usepackage{epsfig,subfigure,psfrag}
\usepackage{dcolumn}
\usepackage{bm}
\usepackage{slashed}
\usepackage{cancel}
\usepackage{color}
\usepackage{tabularx}
\usepackage{appendix}
\preprint{}

\newcommand{\be}{\begin{eqnarray}}
\newcommand{\ee}{\end{eqnarray}}
\newcommand{\bea}{\begin{eqnarray}}
\newcommand{\eea}{\end{eqnarray}}

\newcommand{\gev}{{~\rm GeV}}

\newcommand{\tev}{{~\rm TeV}}

\newcommand{\HT}{\ensuremath{H_T}}

\newcommand{\MET}{\mbox{$E_T\hspace{-0.23in}\not\hspace{0.18in}$}}
\newcommand{\METcaption}{\mbox{$E_T\hspace{-0.21in}\not\hspace{0.18in}$}}

\begin{document}

\title{Hiding Missing Energy in Missing Energy}

\author{Daniele S. M. Alves}
\affiliation{Center for Cosmology and Particle Physics, Department of Physics, New York University, New York, NY 10003}
\affiliation{Department of Physics, Princeton University, Princeton, NJ 08544}

\author{Jia Liu}
\affiliation{Center for Cosmology and Particle Physics, Department of Physics, New York University, New York, NY 10003}

\author{Neal Weiner}
\affiliation{Center for Cosmology and Particle Physics, Department of Physics, New York University, New York, NY 10003}

\date{\today}

\begin{abstract}
Searches for supersymmetry (SUSY) often rely on a combination of hard physics objects (jets, leptons) along with large missing transverse energy to separate New Physics from Standard Model hard processes. We consider a class of ``double-invisible'' SUSY scenarios: where squarks, stops and sbottoms have a three-body decay into {\em two} (rather than one) invisible final-state particles. This occurs naturally when the LSP carries an additional conserved quantum number under which other superpartners are not charged. In these topologies, the available energy is diluted into invisible particles, reducing the {\em observed} missing energy and visible energy. This can lead to sizable changes in the sensitivity of existing searches, dramatically changing the qualitative constraints on superpartners. 
In particular, for $m_{\text{LSP}}\gtrsim 160\gev$, we find no robust constraints from the LHC at any squark mass for any generation, while for lighter LSPs we find significant reductions in constraints. If confirmed by a full reanalysis from the collaborations, such scenarios allow for the possibility of significantly more natural SUSY models. While not realized in the MSSM, such phenomenology occurs naturally in models with mixed sneutrinos, Dirac gauginos and NMSSM-like models.
\end{abstract}

\pacs{Valid PACS appear here}
\maketitle

\section{Introduction}
With the successful operation of the LHC at 7 and 8 TeV energies, experimental results have now probed the energy regime well above the weak scale. While the incredible agreement of the Standard Model is a major success of particle physics, the absence of any clear signs of new physics challenges our basic assumptions about naturalness. 
In particular, it is expected that a top partner should be present to cancel the leading quadratic divergence to the Higgs mass. As a consequence, a hadron collider such as the LHC should be capable of copiously producing such top partners and any other associated colored particles. Specific arguments within supersymmetry for a stable R-parity odd particle, and more generally for a stable T-parity odd particle \cite{Cheng:2003ju} motivate a robust search strategy for jets+missing energy. Such searches have shown no sign of the excesses expected of squarks at several hundred GeV (see, {\it e.g.}, \cite{ATLAS-CONF-2013-024, Aad:2013ija, ATLAS-CONF-2013-048, CMS-PAS-SUS-13-012, ATLAS-CONF-2013-047, Chatrchyan:2013xna}). As a consequence, there is a greater movement to reconsider naturalness entirely \cite{Wells:2004di,Acharya:2008bk,Hall:2011jd,Madrid,Kane:2011kj,SavasFest,Arvanitaki:2012ps,Bhattacherjee:2012ed,ArkaniHamed:2012gw}.

Technically natural models can still be found by restricting the low energy spectrum to the minimal content needed in order to avoid fine-tuning of the electroweak scale (generally stops and Higgsinos with a cutoff) \cite{Papucci:2011wy,Brust:2011tb}. While such scenarios can achieve technical naturalness, they are often {\it ad hoc} in removing other particles from the spectrum (such as unflavored squarks).

The weak scale may still be generically natural, however, if these jets+MET signals are hidden within Standard Model backgrounds. Since large missing transverse energy (MET) is what generally distinguishes these signal events from multijet backgrounds, the simplest possibility is to deform this class of signals by converting MET into visible energy, and hadronic energy in particular. This is realized simply through hadronic R-parity violation, for instance \cite{Carpenter:2007zz,Brust:2011tb}. Detailed questions of flavor violation and baryon number conservation constrain these models \cite{Barbier:2004ez}, but even more pertinent are the constraints from high jet multiplicity searches  \cite{Aad:2013wta, Evans:2013jna} on how well such models hide SUSY.

A second approach is to kinematically suppress missing transverse energy with the presence of nearly degenerate states. This could arise by squeezing the spectrum of squark and bino, for instance, through an accidental degeneracy of the spectrum. Alternatively, ``stealth'' SUSY models \cite{Fan:2011yu,Fan:2012jf} invoke an approximately supersymmetric dark sector to achieve this degeneracy. Both of these approaches attempt to suppress the missing energy by converting as much of the available energy into a visible form. This is successful in suppressing the efficiency of jets + MET searches, but can make other (often dedicated) searches more sensitive, such as \cite{Aad:2011zb,CMS:2012un,Aad:2012zx}.

In this Letter we will consider an alternative possibility - that one can ``dilute"  the final state energy into many invisible particles, and in doing so, obscure signals of New Physics. Momentarily counterintuitive, a brief reflection on the kinematics of the process will make it clear why this suppresses the sensitivity of existing jets+MET searches.

\begin{figure*}[t]
\includegraphics[width=0.45\textwidth]{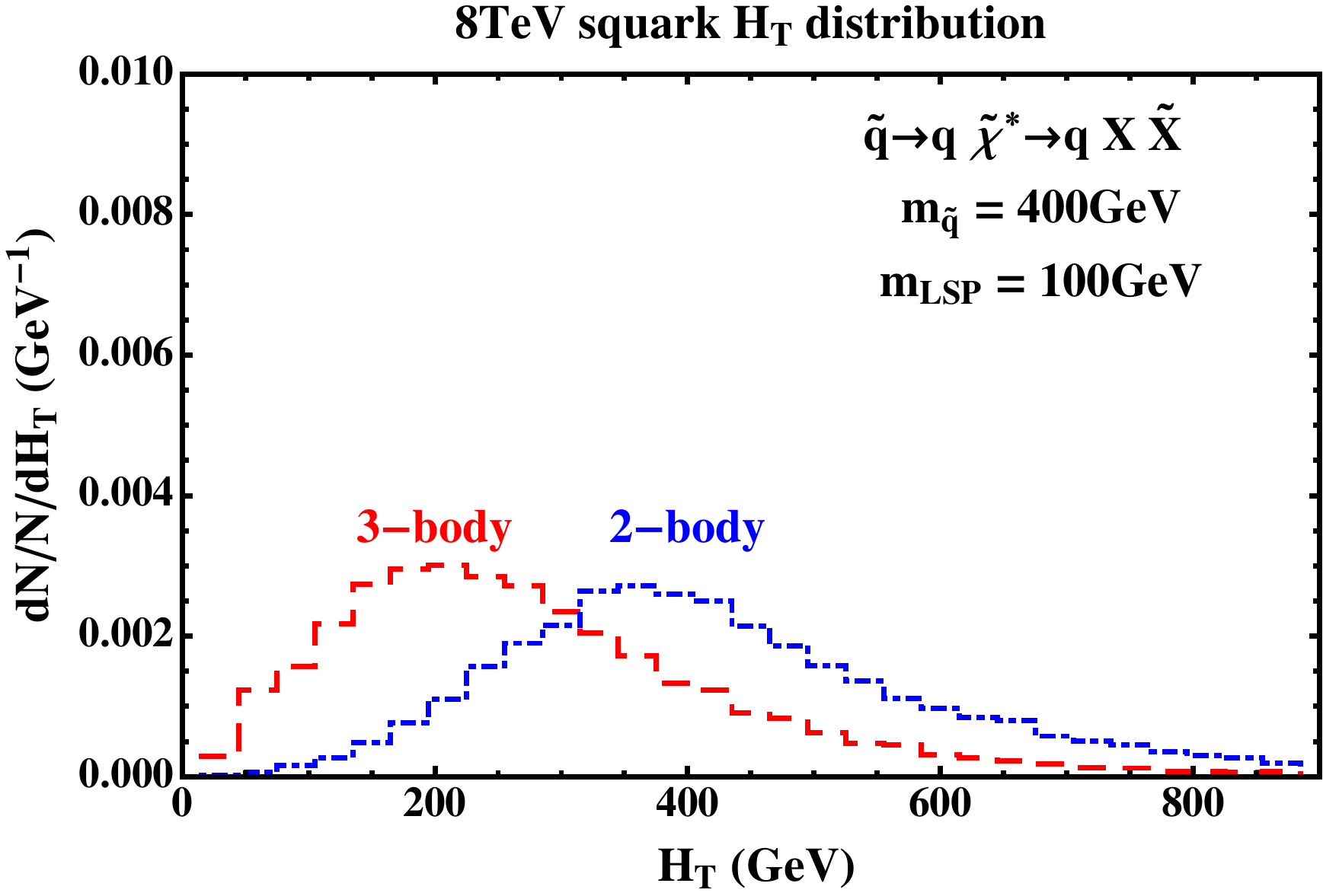} \hskip 0.05\textwidth
\includegraphics[width=0.45\textwidth]{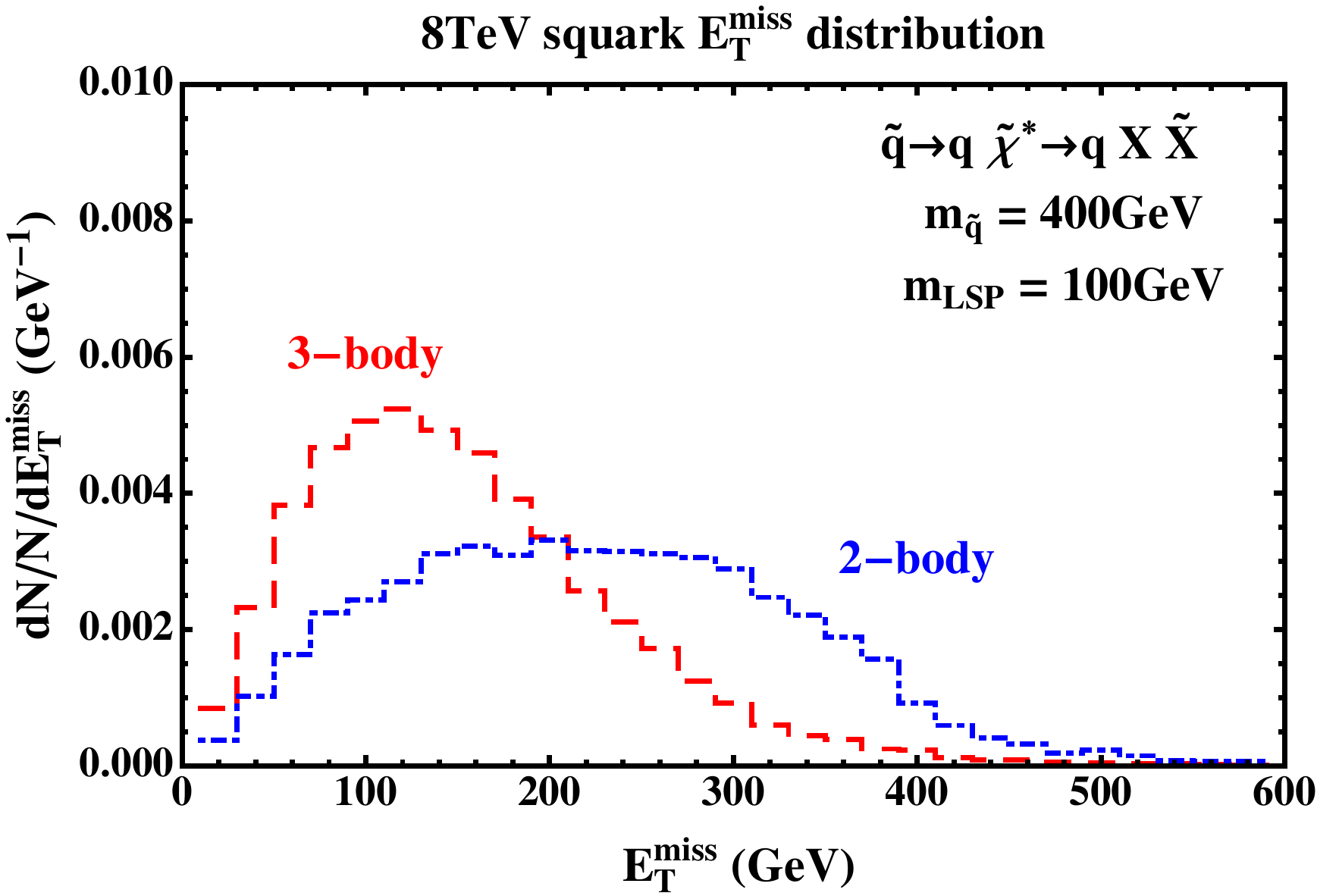} 
\caption{\HT~and $\METcaption$ distributions for squark pair production in the Single-Invisible and Double-Invisible scenarios. In this example, $m_{\tilde q}=400 \gev$ and $m_{\text{LSP}}=100 \gev$. \label{fig:metdist}}
\end{figure*}

\subsection{Hiding Missing Energy in Missing Energy}
The most conventional scenarios in SUSY involve cascades that conclude with a neutralino LSP. In such cases, these cascades generally end with only a single invisible particle - {\it e.g.}, a single squark will cascade to a single R-parity odd neutralino and (mostly) visible energy otherwise. However, this ``single-invisible'' aspect of SUSY is particular to scenarios like the MSSM where the LSP {\em only} carries a single quantum number or parity (in this case R-parity). If the LSP carries a second conserved quantum number not shared by the mother particle, then, to conserve that, there must always be a second stable particle in the cascade (for instance, the R-parity even partner of the LSP). If this particle is invisible, the total amount of missing energy can be increased.

A simple example of this exists already in the MSSM: the sneutrino. Cascades must always conclude with not only the sneutrino, but also an associated lepton. In the case where that lepton is a neutrino, there are two invisible particles in every cascade. Considering the decay of a squark in particular, we can have $\tilde q \rightarrow q \tilde B$ followed by $\tilde B \rightarrow \tilde \nu \nu$. In this case, with an on-shell Bino decaying invisibly, there is no phenomenological difference with simply having a Bino LSP.

In contrast, if the Bino is off-shell, the squark will undergo a 3-body decay, $\tilde q \rightarrow q \tilde \nu \nu$, where the energy is now shared with {\em two} invisible particles. The simplified model that one can consider is one that simply replaces the single invisible decay with a multi-body decay with two invisible particles. We refer to such a scenario and related simplified models as ``double-invisible.'' 

While one might think that increasing the multiplicity of invisible particles in the final state would increase the sensitivity of jets+MET searches, the opposite is actually true. This is because the extra invisible states dilute the energy of the visible particles. Since MET ($\MET$) is a vector-sum of visible energy, the increase in missing (scalar-sum) energy leads to a decrease in missing (vector-sum) energy. 
We can see an example of this in Fig.~\ref{fig:metdist}. 
These changes naturally have a significant impact on SUSY searches.


\section{Experimental Sensitivity on Double-Invisible Simplified Models}\label{sec:limits}
Generically, SUSY searches for colored superpartners are optimized for standard (single-invisible) MSSM decays. That typically entails hard cuts on missing energy, hadronic energy and leading jets' transverse momenta. Such cuts substantially reduce backgrounds without compromising sensitivity to standard topologies. However, hard requirements on kinematics can lead to a significant reduction of signal efficiency for double-invisible topologies, as suggested by the distributions on Fig.~\ref{fig:metdist}.

In this section, we will attempt to recast \cite{Cranmer:2010hk} the limits from ATLAS and CMS SUSY searches to the double-invisible scenario. As we shall see, they are significantly weakened, by our estimates by almost an order of magnitude in cross section at times.

Before we lay out our goals, we should emphasize that our limits should not be taken as precise limits, but as our best current estimates, and as motivations for the experiments to properly recast these limits themselves. Secondly, we would argue that these limits motivate new analyses, more optimized for these kinematics. As 13 TeV data may be more challenging to apply to these low masses, such analyses should be a high priority prior to the next LHC run.

\begin{figure*}[t]
\includegraphics[width=0.51\textwidth]{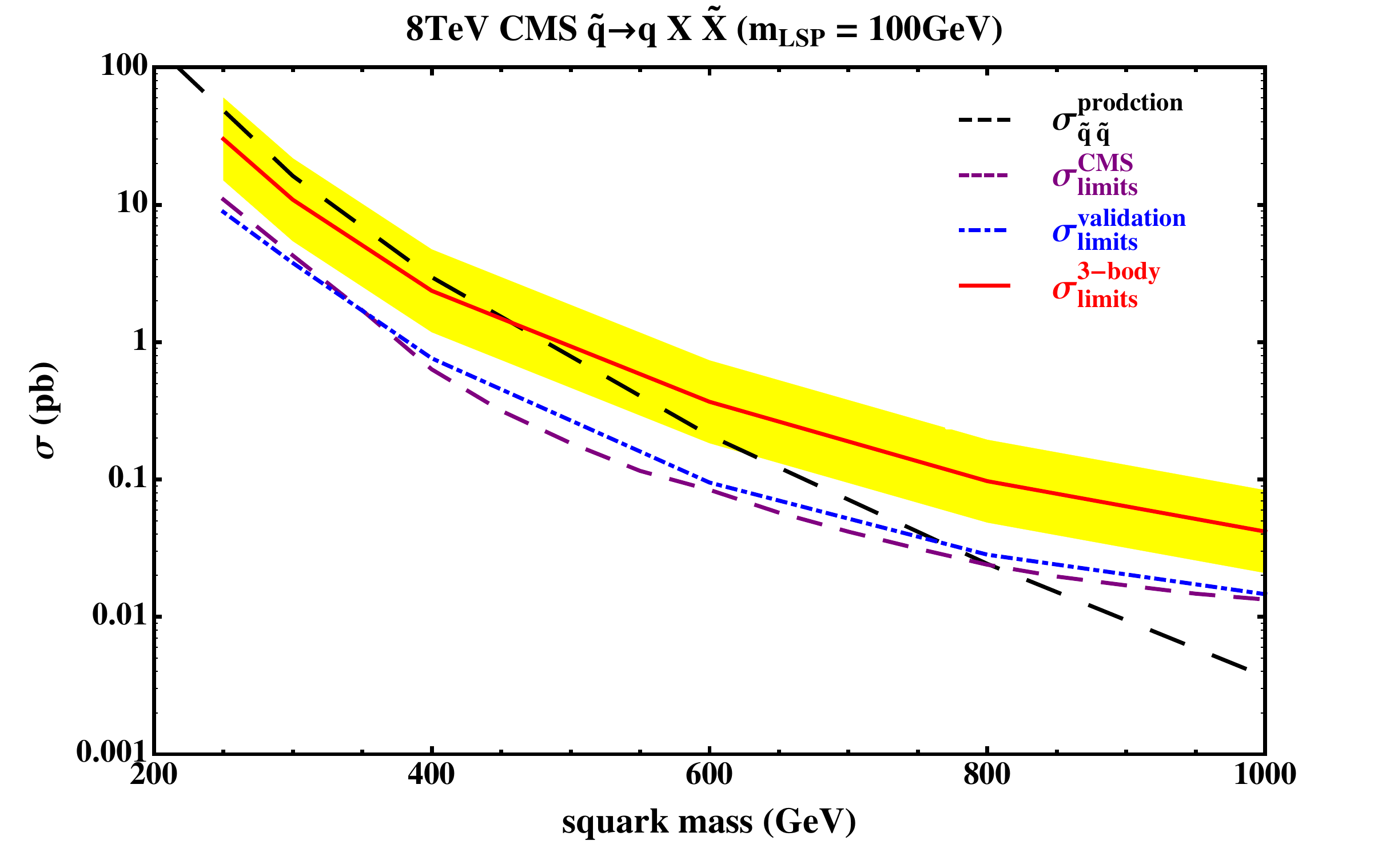} \hskip 0.05\textwidth
\includegraphics[width=0.40\textwidth]{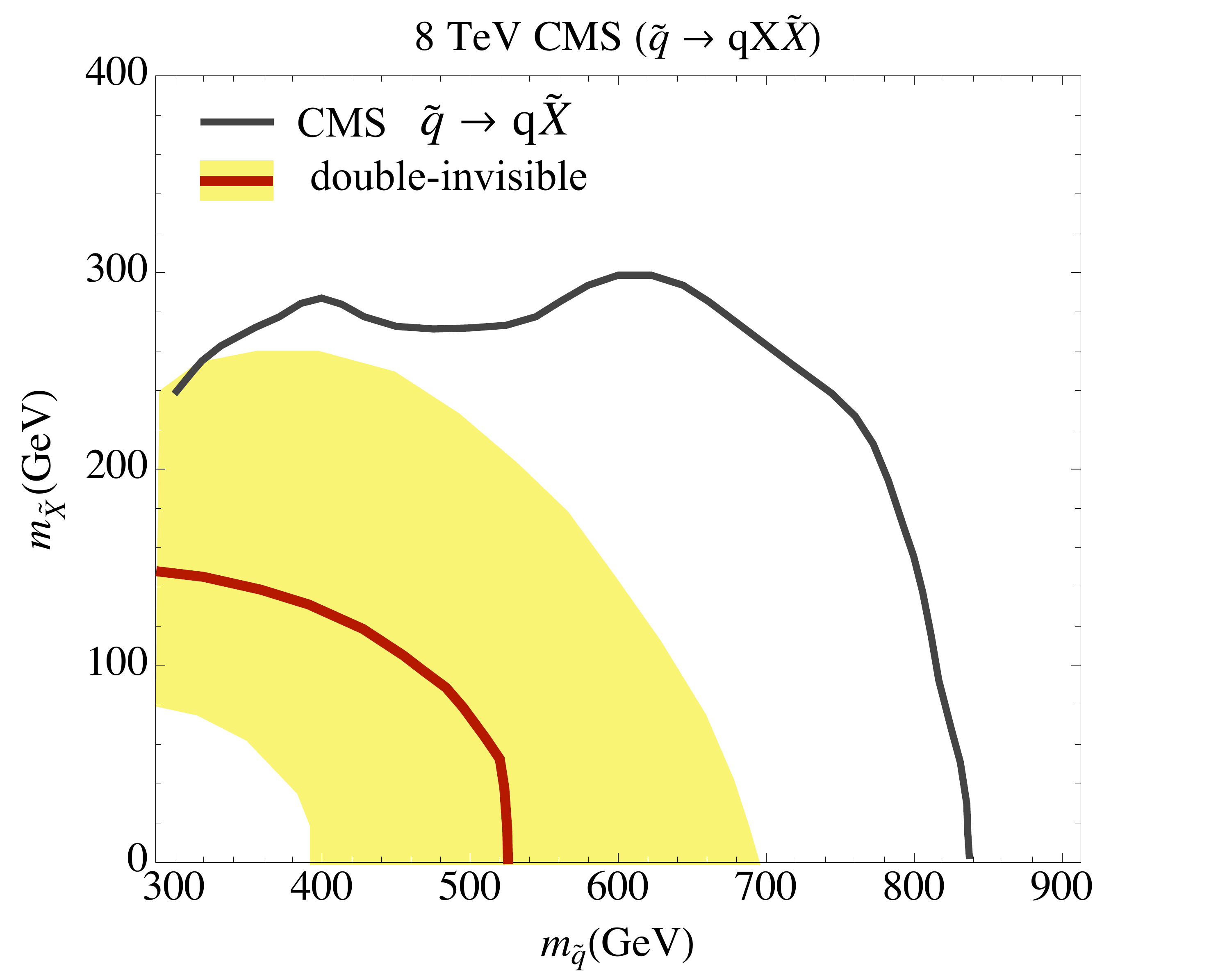} 
\caption{CMS constraints on degenerate 1st and 2nd generation squarks for single- and double-invisible SUSY scenarios. ATLAS constraints are weaker for this topology. The (shaded) yellow band corresponds to an {\it ad hoc} factor of two uncertainty in our estimated limits. \label{fig:lightsquarks}}
\end{figure*}

We generate Monte Carlo events for double-invisible simplified models and survey their constraints from relevant ATLAS and CMS searches.
In order to validate our simulation and calculation of the experimental efficiencies, we first attempt to reproduce the experimental limits quoted by the searches. We only present our estimated limits for analyses we were able to validate, i.e., whose results we were able to reproduce to within a factor of two.

We simulate pair-production of colored superpartners in Madgraph 5 \cite{Alwall:2011uj}, which are decayed, showered and hadronized in Pythia 6 \cite{Sjostrand:2006za}. For a crude simulation of detector response, we use PGS4 \cite{pgs4}. For searches requiring b-jets, we have modified PGS's b-tagging efficiency as a function of the b-jet's transverse momentum and rapidity in order to more closely match the working point used by the relevant searches.

For squarks and gluinos, we validated and recast the searches in \cite{ATLAS-CONF-2013-047, CMS-PAS-SUS-13-012}. The validated and recast analysis for third generation squarks were \cite{Aad:2013ija, ATLAS-CONF-2013-024, ATLAS-CONF-2013-048, Chatrchyan:2013xna}. Other potentially relevant searches will not be discussed in this note either because we have found that they were not competitive with the analyses listed above, or because we were not able to validate their limits to a satisfactory degree. Instances of the former category are $\alpha_T$, razor and monojet searches. We expect a lower sensitivity of the CMS $\alpha_T$ analysis in \cite{Chatrchyan:2013lya} due to its lower luminosity (11.7 $\text{fb}^{-1}$) and hard requirements on the transverse energy of the two leading jets ($E_T^{j_1,j_2}\geq 100 \gev$).  The CMS razor analyses, at the time of writing of this letter, have not been updated with the 8 TeV data. Even though we might expect non-trivial 7 TeV razor limits to our scenario, we do not expect that they will be stronger than other 8 TeV hadronic analyses with higher energy and four to five times more integrated luminosity. As for the monojet analyses, ATLAS has a dedicated search for compressed stops decaying to a charm quark and a neutralino \cite{ATLAS-CONF-2013-068}, excluding the very compressed region with $m_{\tilde{t}} \lesssim 230 \gev$. Their limits can be straightforwardly recast to eight compressed squarks of the 1st and 2nd generations, being roughly $m_{\tilde{q}} \gtrsim 360 \gev$. We expect this search to have a reduced efficiency on non-compressed double-invisible topologies, and therefore can be ignored for our purposes for not being competitive with the CMS limits from \cite{CMS-PAS-SUS-13-012}. The second category of searches not included in this study, {\it i.e.}, those we cannot validate, spans searches that use multivariate analyses, neural networks, boosted decision trees, etc., for which we do not have enough information or tools to reproduce.

Fig.~\ref{fig:lightsquarks} shows our recast limits from \cite{CMS-PAS-SUS-13-012} on degenerate 1st and 2nd generation squarks with 3-body decay $\tilde{q}\rightarrow q X\tilde{X} $, where $X$ and $\tilde{X}$ are invisible and $m_X=0$. Gluinos are assumed to be decoupled. In the left plot, we set $m_{\tilde X} = 100~\gev$ and show the limit on the production cross section as a function of the squarks' mass (red line). The shaded yellow band corresponds to an {\it ad hoc} factor of two uncertainty in our estimates. We also show the reference NLO-QCD production cross section (black line) computed with Prospino 2 \cite{Beenakker:1996ed}, the official CMS limits on the standard two-body topology (purple line) and our validation of the CMS limits (blue line). One can see that for most of the squark mass range, the cross section limits we find on the double-invisible topologies are reduced by roughly a factor of 5 relative to their single-invisible counterparts. Squark mass limits are weakened from $m_{\tilde q} \lesssim 800~\gev$ to $m_{\tilde q} \lesssim 450~\gev$ assuming $m_{\tilde X} = 100~\gev$, and  disappear for $m_{\tilde X} \gtrsim 160~\gev$, as shown on the plot on the right, which contrasts double and single-invisible constraints in the $m_{\tilde q} - m_{\tilde X}$ plane. Interestingly, our recast of the ATLAS jets+MET search \cite{ATLAS-CONF-2013-047} on this topology yielded no constraint on squark masses, regardless of $m_{\tilde X}$. That can be explained by the tight cuts applied to the event selection, in particular to the leading jet transverse momentum ($p_T^{j_1}\geq 130 \gev$).

\begin{figure*}
\includegraphics[width=0.47\textwidth]{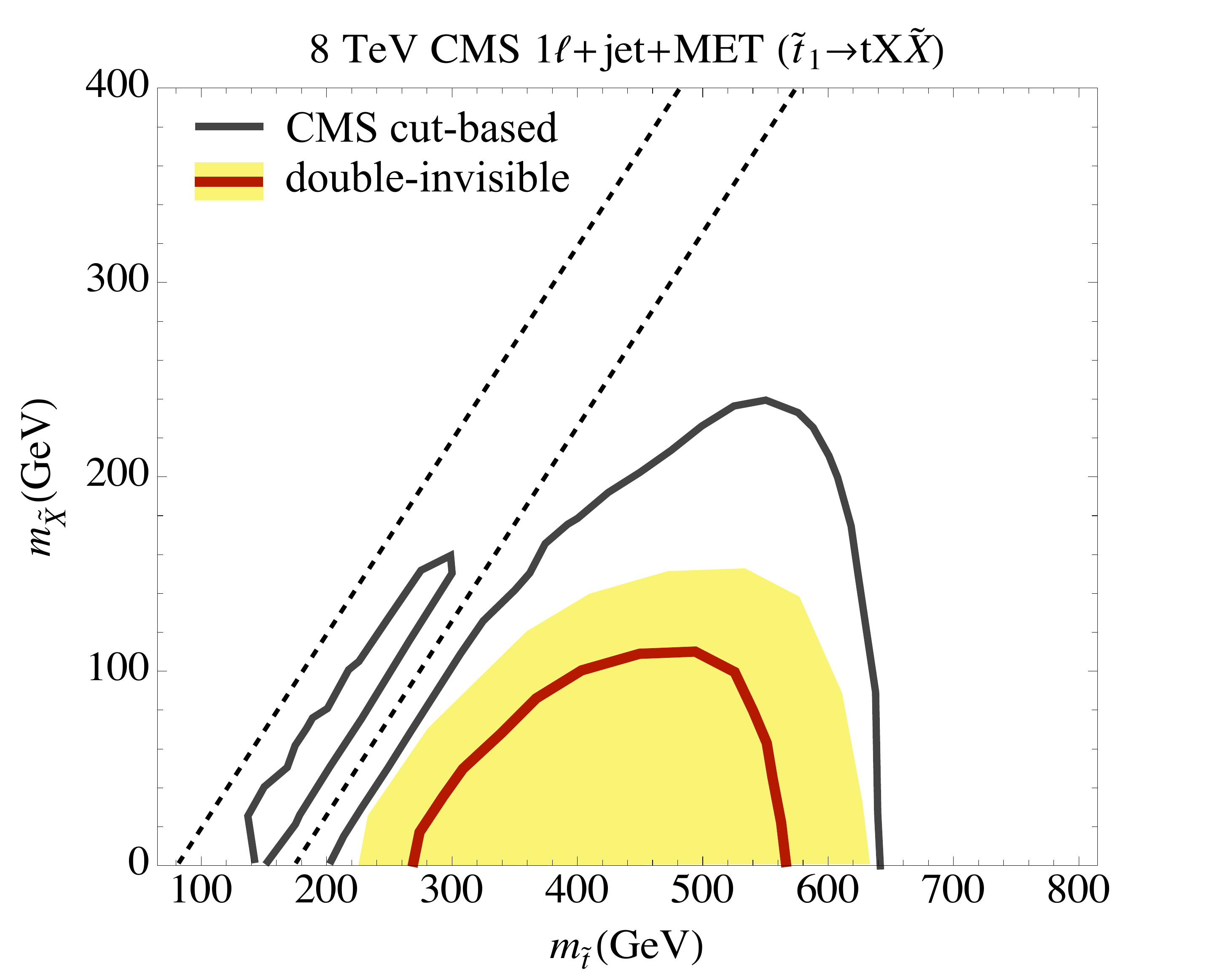} \hskip 0.05\textwidth
\includegraphics[width=0.46\textwidth]{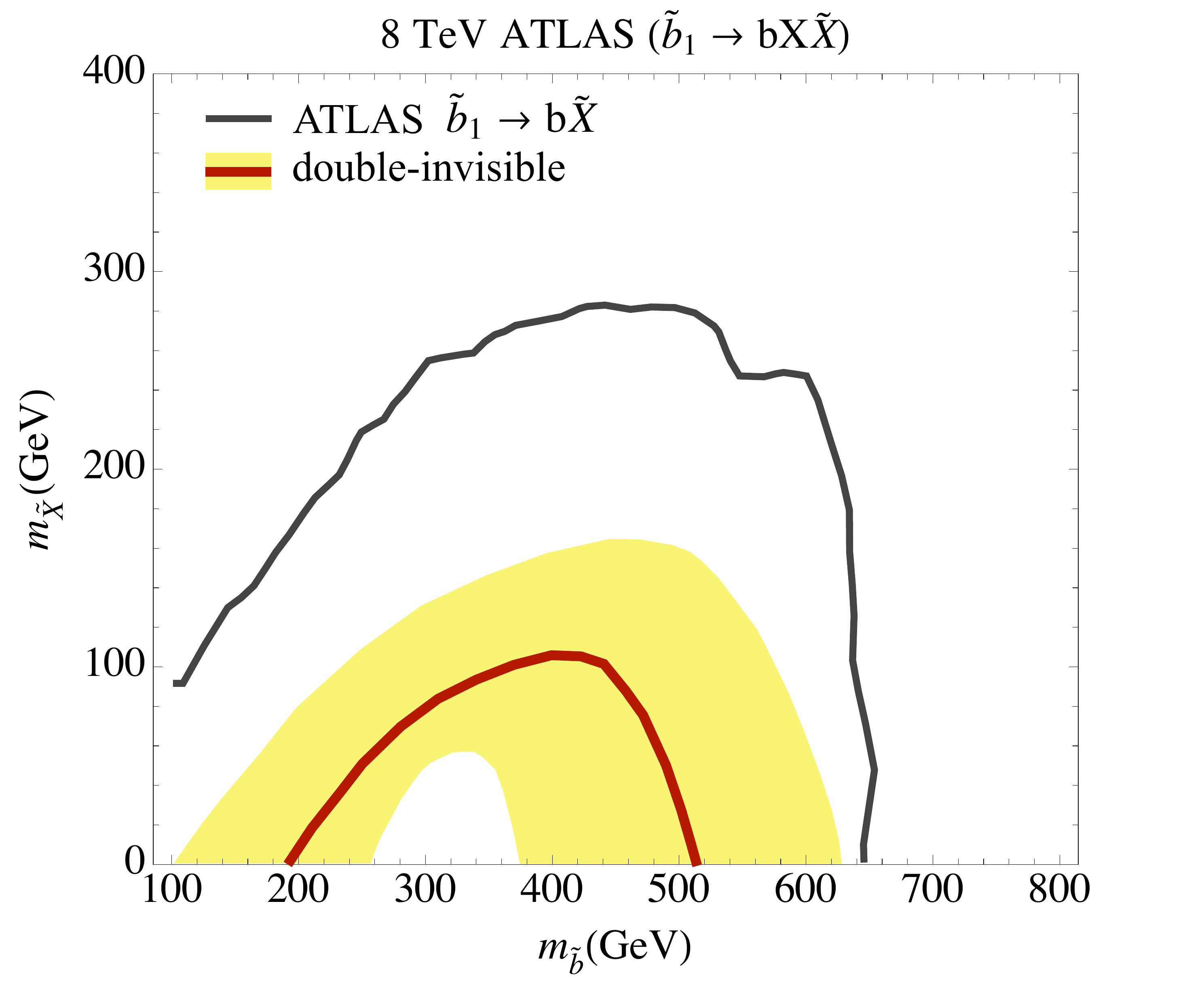} 
\caption{Limits on 3rd generation squarks for the single- and double-invisible SUSY scenarios. As in Fig. \ref{fig:lightsquarks}, the (shaded) yellow band corresponds to an {\it ad hoc} factor of two uncertainty in our estimated limits.  \label{fig:stopsandsbottoms}}
\end{figure*}

Fig.~\ref{fig:stopsandsbottoms} shows our estimated limits for 3rd generation squarks in the $m_{\tilde{b}/\tilde{t}} - m_{\tilde X}$ plane. Again we assume $m_X=0$ for the purpose of illustration and add an {\it ad hoc} factor of two uncertainty in our estimates, delimited by the yellow region. We only display the limits from \cite{Chatrchyan:2013xna, Aad:2013ija}, which are the most sensitive to the topologies $\tilde{t}\rightarrow t X\tilde{X} $ and $\tilde{b}\rightarrow b X\tilde{X} $ (constraints from other 3rd generation searches are shown in the Appendix). These plots again suggest that bounds on stops and sbottoms are substantially reduced for double-invisible topologies, even disappearing for $m_{\tilde X} \gtrsim 120~\gev$.

As previously mentioned, the limits just discussed assume decoupled gluinos. If gluinos are kinematically accessible, one has to consider additional colored production, such as $p p \rightarrow \tilde g \tilde g$, $\tilde g \tilde{q}^{(*)}$ and $\tilde q \tilde q$ (the later being enhanced via t-channel gluino). That can substantially increase the constraints on squarks, for instance $m_{\tilde q}\gtrsim 1380~\gev$ for $m_{\tilde q} = 0.96\times m_{\tilde g}$. For $m_{\tilde q}= 500\gev$, the gluino must be heavier than $\sim 2.5-3 \text{ TeV}$. Such a separation could be natural if gluino and squark masses are generated at a low scale, with $m_{\tilde q}^2$ two-loop suppressed relative to $M_{\tilde g}$ (as occurs with Dirac gauginos \cite{Fox:2002bu}).

\section{Model realizations of Double-Invisible SUSY}\label{sec:models}
Model realizations of double-invisible SUSY are straightforward (but not trivial) to construct. There are two essential elements for the model: first, the LSP $\tilde X$ must carry some additional charge or parity (not shared by other superpartners) so that it is always accompanied by an additional particle $X$ carrying that same charge or parity. Moreover, this additional particle must be neutral. \footnote{In the MSSM, only one final state fits these criteria, namely $\tilde \nu$ and $\nu$, with the additional charge being lepton number. While the 3-body decay $\tilde q \rightarrow q \tilde \nu \nu$ can be accommodated in a mixed sneutrino scenario \cite{ArkaniHamed:2000bq}, it would hardly be the dominant one if the charged slepton decay channel is open.}

Having the appropriate final state is not enough, obviously, as the 3-body decay $\tilde q \rightarrow q X \tilde X$ must be the dominant decay mode. If the {\em only} R-parity-odd and kinematically open channel is $X \tilde X$, then the double-invisible phenomenology is realized fairly trivially. However, this dictates a somewhat specific class of spectra, with squarks the next-to-lightest sparticles. We would be interested in exploring whether models can exist with additional light sparticles but retaining the double-invisible phenomenology.

It is fairly clear that for two-body decays to be suppressed, the gauginos must be heavier than the squarks. As discussed in Sec.~\ref{sec:limits}, for light squarks ($m_{\tilde q}\sim 500 \gev$), the gluino must satisfy $m_{\tilde g} \gtrsim 2.5 - 3 \tev$. Such a separation between squarks and gluinos is most natural in the context of Dirac gauginos, where the loop corrections to the squark masses squared are ``supersoft'', or finite to all orders \cite{Fox:2002bu}. Moreover, in this scenario the gluino t-channel contribution to squark pair-production is suppressed \cite{Choi:2008pi, Kribs:2012gx}, further reducing limits on squark production. Because Dirac gauginos seem to provide the natural basic framework in which such phenomenology is viable, we shall focus our model building efforts there.

We add to the MSSM Lagrangian terms
\be
W=\frac{1}{M_{med}}W_Y^\alpha W_\alpha' S + y S X \bar X + m X \bar X,
\ee
where $\langle W_\alpha' \rangle = \theta D$ is an effective D-term spurion (which may arise from the D-term of a hidden sector $U(1)'$ or from a composite vector $\langle \bar D^2 D^\alpha X^\dagger X \rangle = \theta F^2$). 
We assume the first term provides the dominant contribution to the Bino mass. Note that while we have included a mass term for $X$, the vev for $S$ induced after EWSB will generate a small $X$ mass in the absence of an explicit mass term. 
Note that we use $ \sim$ to denote the R-parity odd state here, but there is a choice whether that is the scalar or fermion state (or, equivalently, whether to expand the definition of R-parity to include the $X$-charge).

Assuming sleptons are kinematically accessible, the partial width for leptonic decays scales as $\Gamma_{\tilde q \rightarrow q l \tilde l} \propto g_Y^4 m_{\tilde q}^5/m_{\tilde B}^4$, while the double-invisible decay scales as $\Gamma_{\tilde q \rightarrow q X \tilde X} \propto g_Y^2 y^2 m_{\tilde q}^3/m_{\tilde B}^2$. The different scaling is due to the fact that the Dirac mass insertion on the Bino propagator flips to a right-handed state that has no couplings to SM leptons \cite{Kribs:2007ac}. Consequently, the branching ratio to charged leptons will fall as $\text{Br}({\tilde q \rightarrow q l\tilde l}) \sim (g_Y^2 m_{\tilde q}^2)/(y^2 m_{\tilde B}^2)$ and will be sufficiently suppressed for $m_{\tilde B}\gtrsim\mathcal{O}(\text{TeV})$ and $y \sim \mathcal{O}(1)$, allowing the double-invisible phenomenology to dominate.

\subsection{Displaced Scenarios}
If squarks are the next-to-lightest R-parity odd superpartners ($\tilde X$ being the LSP), another intriguing possibility arises, namely that of displaced decays. Since the decay arises from a higher dimension operator, displaced decays can be quite natural.

Rather than decaying the squarks through the Bino portal as above, one can consider the Higgs portal, by adding to the MSSM Lagrangian the terms
\be
W=\mu H_u H_d + \lambda S H_u H_d + m S^2 + y S X \bar X + m X \bar X.
\ee
The decay $\tilde q \rightarrow q X \tilde X$ will proceed either via mixing with the Higgsino (and thus with an amplitude proportional to $y$, $\lambda$, and the fermion's Yukawa, $y_f$) or via the Bino through its Higgsino mixing, and thus proportional to $y$, $\lambda$ and $m_Z/m_{\tilde B}$. This raises the possibility that the squark decay will be displaced. The phenomenology will be similar to that in ``mini-split'' scenarios \cite{Arvanitaki:2012ps,ArkaniHamed:2012gw}, where the gluino decays through a dipole operator to a gluon and a neutralino. Here, however, such signals arise at a lower energy scale, and the cross section magnitude is set by squark pair-production, rather than gluino pair-production.

\subsection{N-Invisible SUSY}
While we have focused so far on double-invisible SUSY, it is straightforward to extend the scenario to a multibody decay with $N$ invisible final-state particles. As multibody decays are inevitably from higher dimension operators, the displaced scenario is much more likely here. Putting that aside for the moment (assuming the intermediate states are sufficiently light to allow prompt decays), we can consider a modification to the above model with the additional terms 
\be
W\supset X Y^2 + m Y \bar Y.
\ee
If the decay $\tilde q \rightarrow  q X \bar X$ is kinematically forbidden because, say, the scalar $X$ is too heavy, then the decay $\tilde q \rightarrow  q \bar X Y Y$ will be the only allowed one, realizing a ``triple-invisible'' scenario.

Admittedly, this particular model realization is somewhat contrived, and adding additional fields to achieve four and five invisible particles in the final state may be more so. Nonetheless, these are still logical possibilities and warrant a recast of existing analyses, if not dedicated analyses.

\section{Conclusions}
The successful run of the LHC at 7 and 8 TeV has significantly constrained a large number of scenarios for physics beyond the Standard Model. In particular, most conventional SUSY models are tightly constrained unless the majority of colored particles are above $\mathcal{O}(\text{TeV})$. 

Such limits can be dramatically alleviated in ``double-invisible'' supersymmetric scenarios, in which squarks 3-body decay into a quark and two invisible particles, rather than a single neutralino. Such scenarios are natural if the LSP carries a new conserved quantum number (or parity) such that it must be produced with an R-parity even partner.

In those scenarios, the total energy carried away by the invisible particles is increased, diluting the visible energy in the final state. While a (naive) paradox, this increased invisible energy decreases the measured missing energy, thus lowering the sensitivity of existing searches to squarks decaying double-invisibly. In particular, our recasts of the existing ATLAS and CMS searches indicate that for $m_{\rm LSP} \gtrsim 160 \gev$ and $m_{\tilde g} \gtrsim 3 \tev$, (unflavored) squarks, sbottoms and stops lack any robust LHC constraints (in large contrast with the strongly constrained parameter space of their single-invisible counterparts). Non-trivial limits still hold for lighter LSP masses, $m_{\rm LSP} \lesssim 160 \gev$, though substantially reduced. This goes counter to the conventional wisdom that colored particles decaying into jets+MET are tightly constrained, unless a kinematical tuning suppresses the missing energy. At a minimum, this warrants a proper analysis of these scenarios by the ATLAS and CMS collaborations and, should those be as weak as our study suggests, dedicated searches should be performed taking into account the modified kinematics. We emphasize that the two limiting cases ($m_{X}=0$ and $m_{X}=m_{\tilde X}$) have no additional parameters beyond the usual simplified models of squarks and neutralinos, making a thorough study viable.

Models with ``multi-invisible" phenomenology can be constructed easily, but in particular find a natural home with Dirac gauginos. While the Dirac gaugino framework has its own issues \cite{Fox:2002bu,Amigo:2008rc,Arvanitaki:2013yja,Csaki:2013fla}, the possibility of light squarks and a genuinely ``natural'' weak scale remains, motivating further study.

Regardless of whether the phenomenology presented in this Letter is realized in nature, it highlights the importance of not assuming that the few-hundred GeV scale has been thoroughly explored for colored particles. Especially as the LHC moves on to even higher energies, it is essential to remain critical of existing searches to make sure some subtlety has not caused us to miss New Physics under our noses.

\begin{acknowledgments}

\end{acknowledgments}
The authors thank R. D' Agnolo, T. Cohen, J. Evans, and J. Ruderman for useful discussions. We also thank Eva Halkiadakis, Christian Sander and Seema Sharma for patiently answering our detailed questions and making efficiency files available on their TWiki. DA is supported by NSF-PHY-0969510 (the LHC Theory Initiative). NW and DA are supported by the NSF under grants PHY-0947827 and PHY-1316753. 

\vspace{0.2in}
\begin{appendix}
\section{Extra Constraints}

In this Appendix, we provice further contraints on double-invisible topologies from the searches we have validated. Throughout the plots we assume that the R-parity-even state is massless. The shaded yellow area denotes a factor of two uncertainty in our estimated cross section limits.

\begin{figure*}
\includegraphics[width=0.47\textwidth]{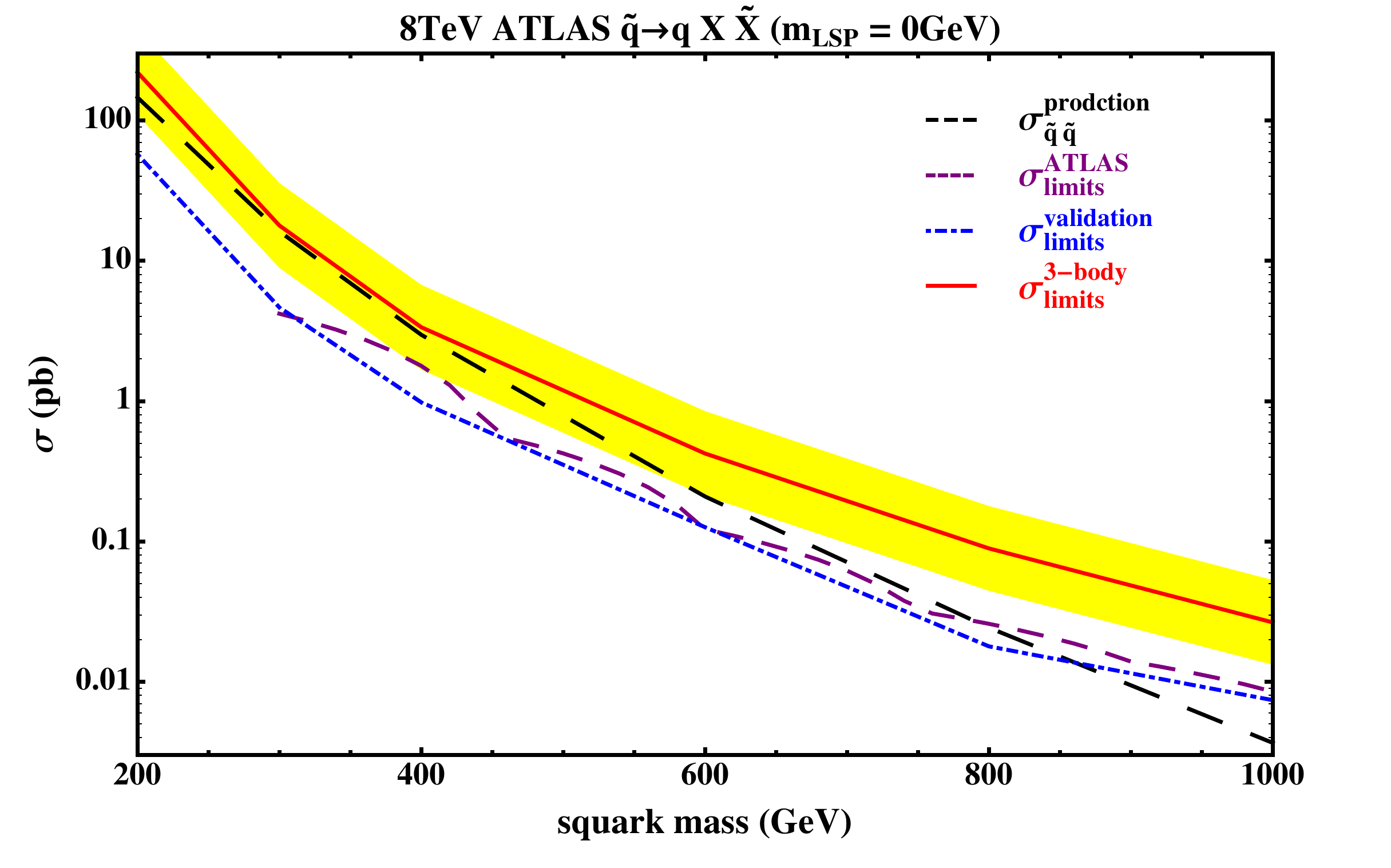}\hskip 0.05\textwidth
\includegraphics[width=0.47\textwidth]{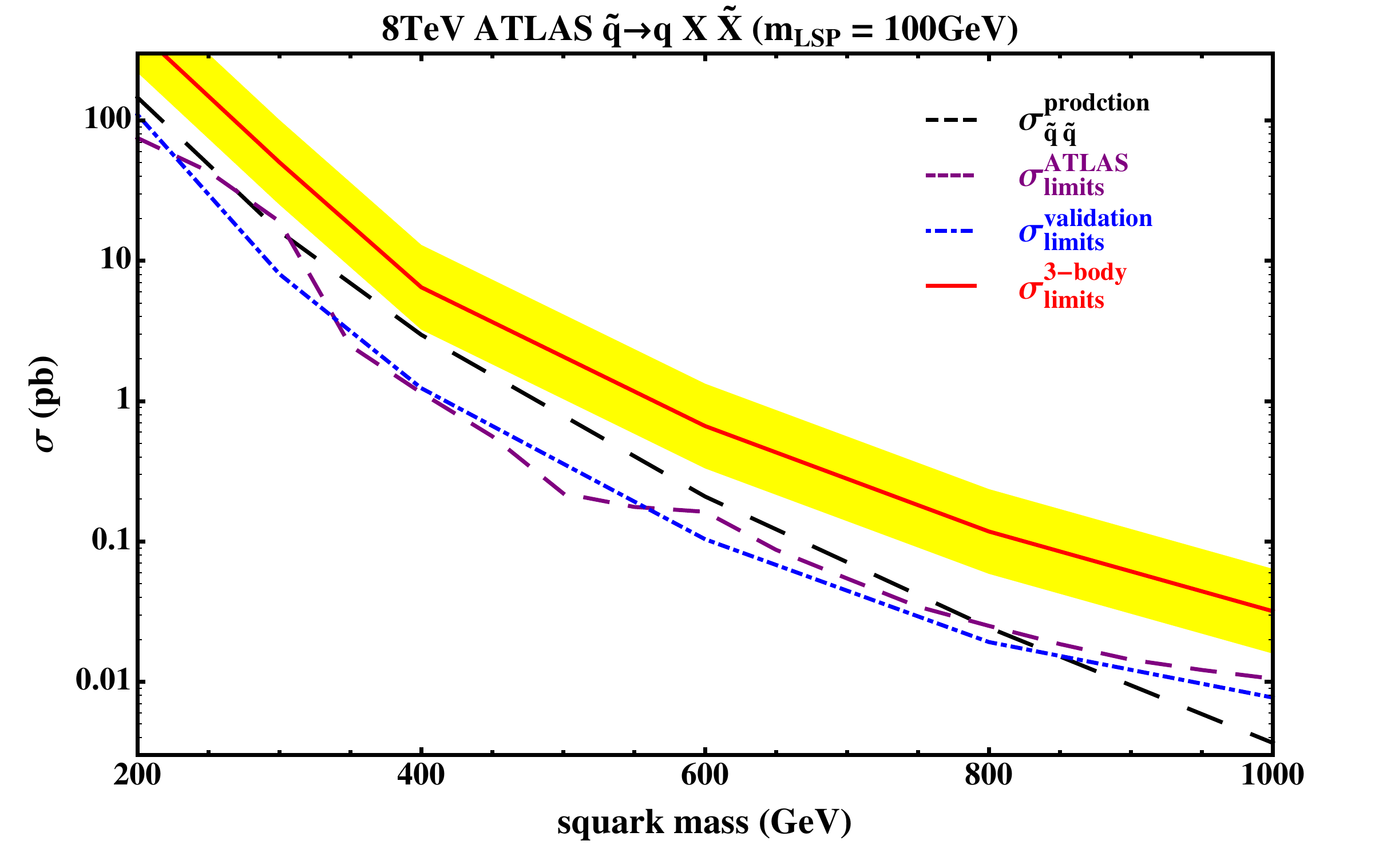}
\caption{ATLAS \cite{ATLAS-CONF-2013-047} limits on 1st and 2nd generation degenerate squarks with gluinos decoupled, with $m_{\tilde X}=0$ (left) and $m_{\tilde X}=100 \gev$ (right). As mentioned in Sec.~\ref{sec:limits}, ATLAS does not place any robust limits on squarks decaying double-invisibly due to their hard selection cuts.}
\end{figure*}

\begin{figure*}
\includegraphics[width=0.47\textwidth]{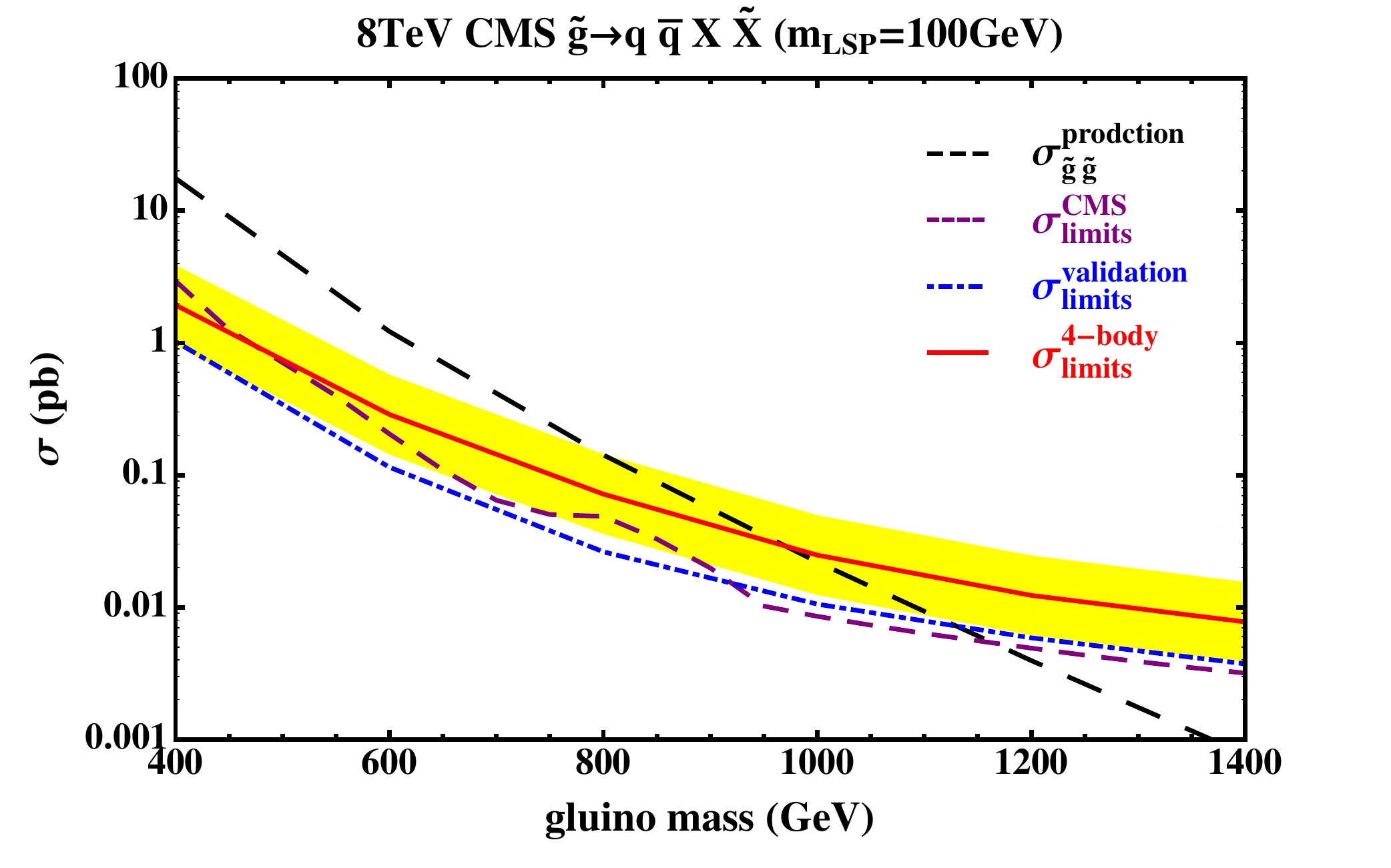} \hskip 0.05\textwidth
\includegraphics[width=0.47\textwidth]{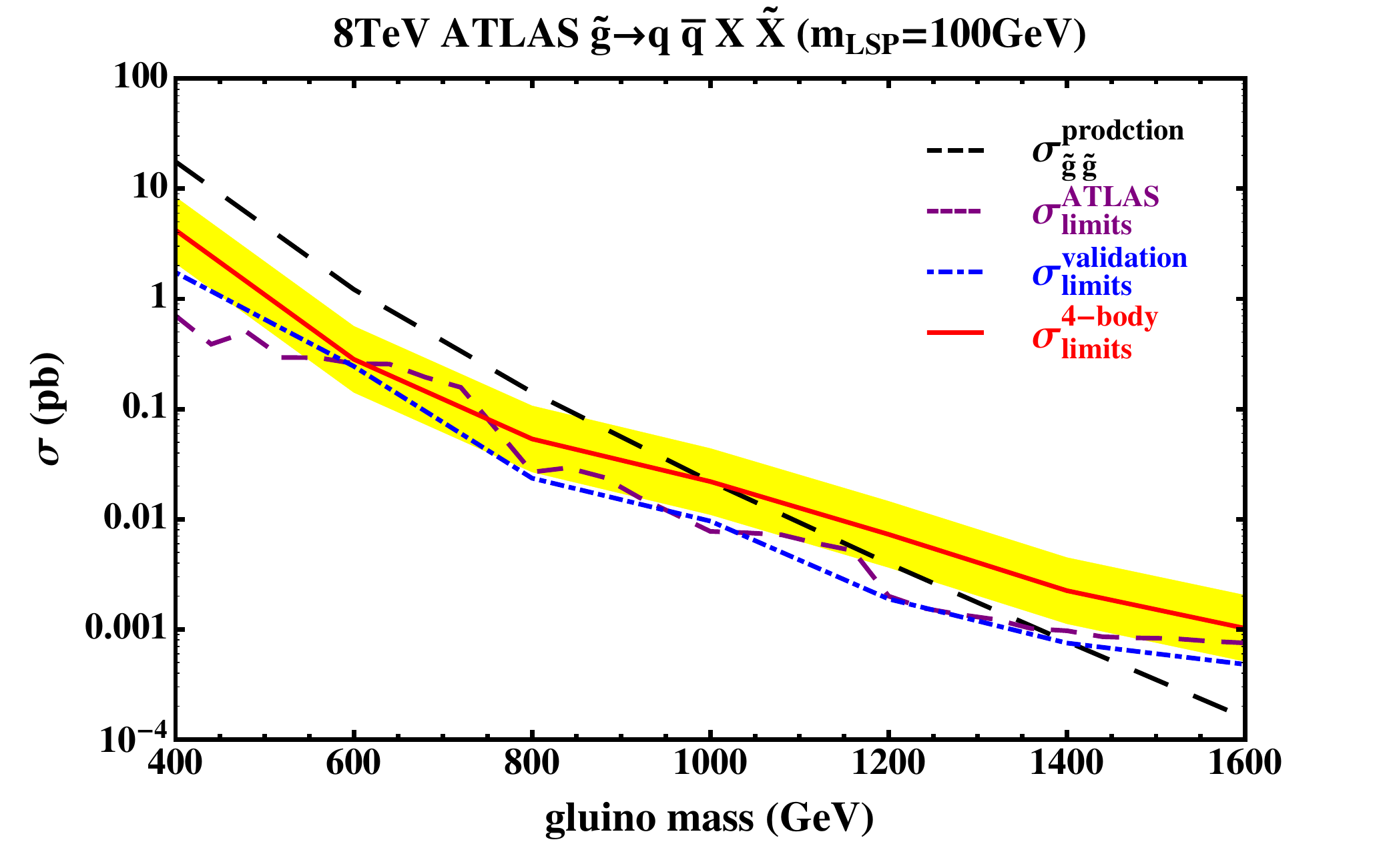} 
\caption{CMS \cite{CMS-PAS-SUS-13-012} and ATLAS \cite{ATLAS-CONF-2013-047} constraints on gluino pair-production with decoupled squarks. In the double-invisible scenario, the gluino undergoes a 4-body decay $\tilde g \rightarrow q q \tilde X X$.}
\end{figure*}

\begin{figure*}
 \includegraphics[width=0.47\textwidth]{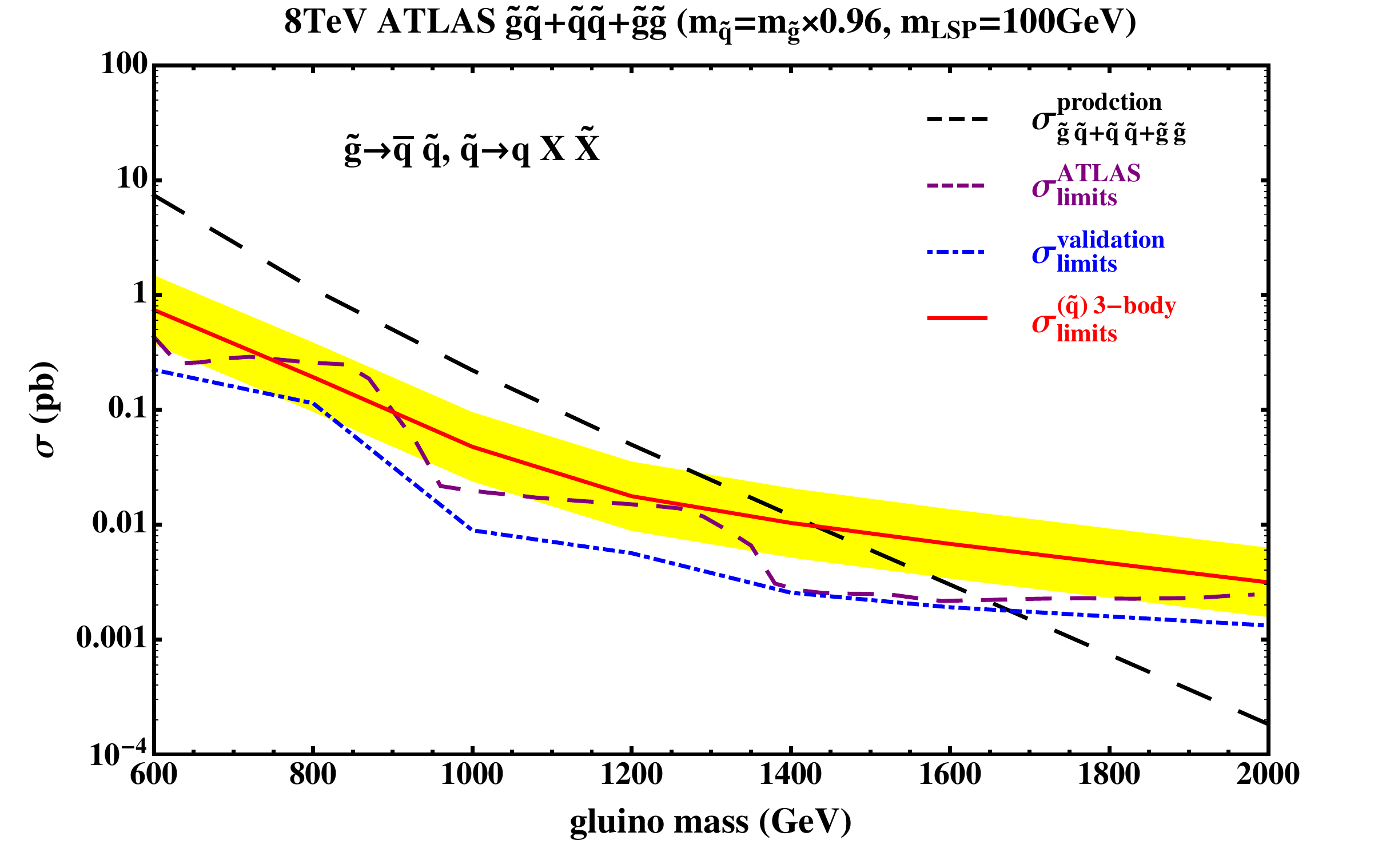} \hskip 0.05\textwidth
 \includegraphics[width=0.47\textwidth]{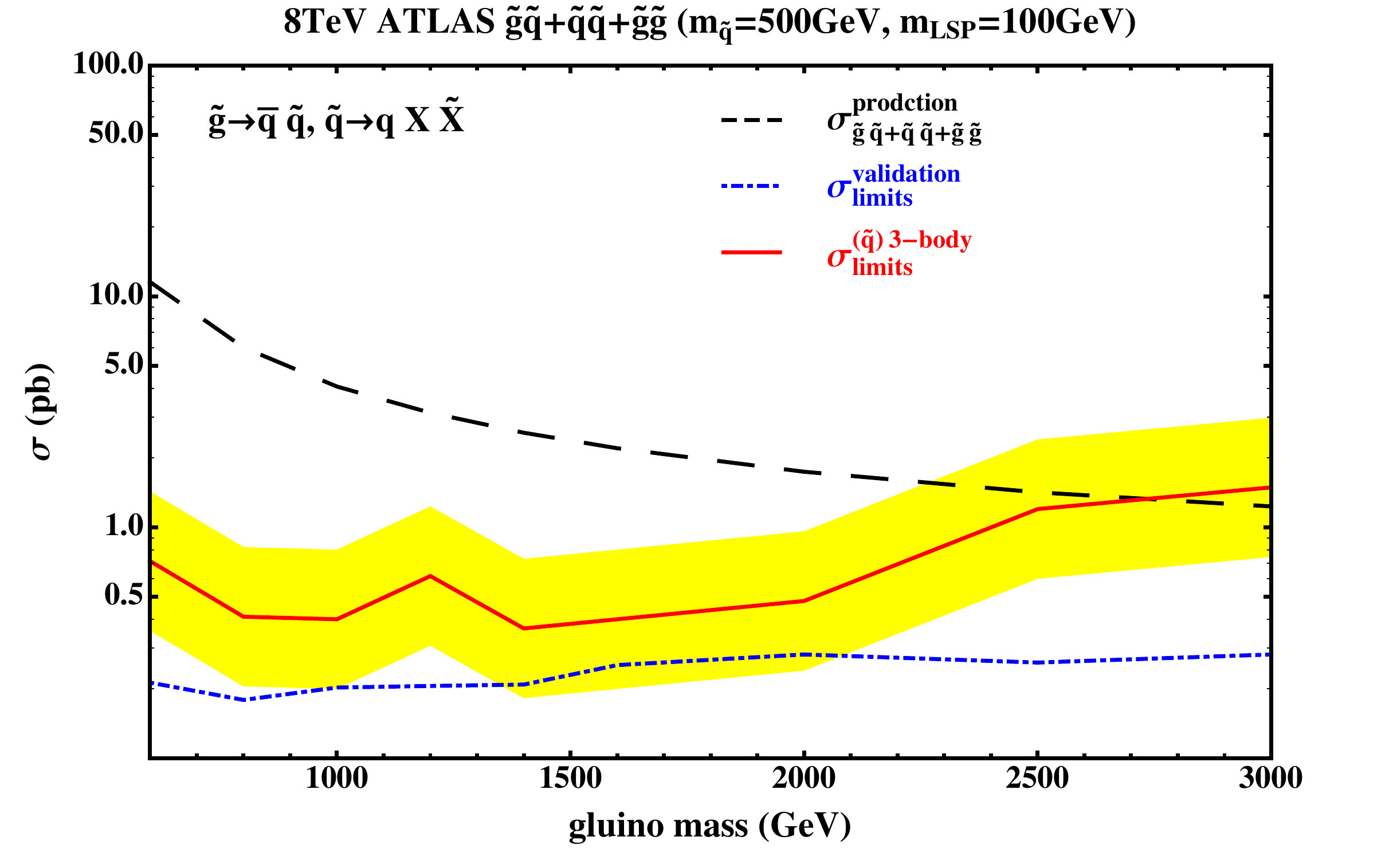}
\caption{ATLAS \cite{ATLAS-CONF-2013-047} constraints on colored production with degenerate 1st and 2nd generation squarks and gluinos kinematically accessible. ({\it Left}) A fixed mass ratio $m_{\tilde q}= 0.96\times m_{\tilde g}$ is assumed. ({\it Right}) Squark masses are fixed to $m_{\tilde q}=500 \gev$. In this case, ATLAS does not provide official cross section limits in the single-invisible topology.}
\end{figure*}


\begin{figure*}
\includegraphics[width=0.47\textwidth]{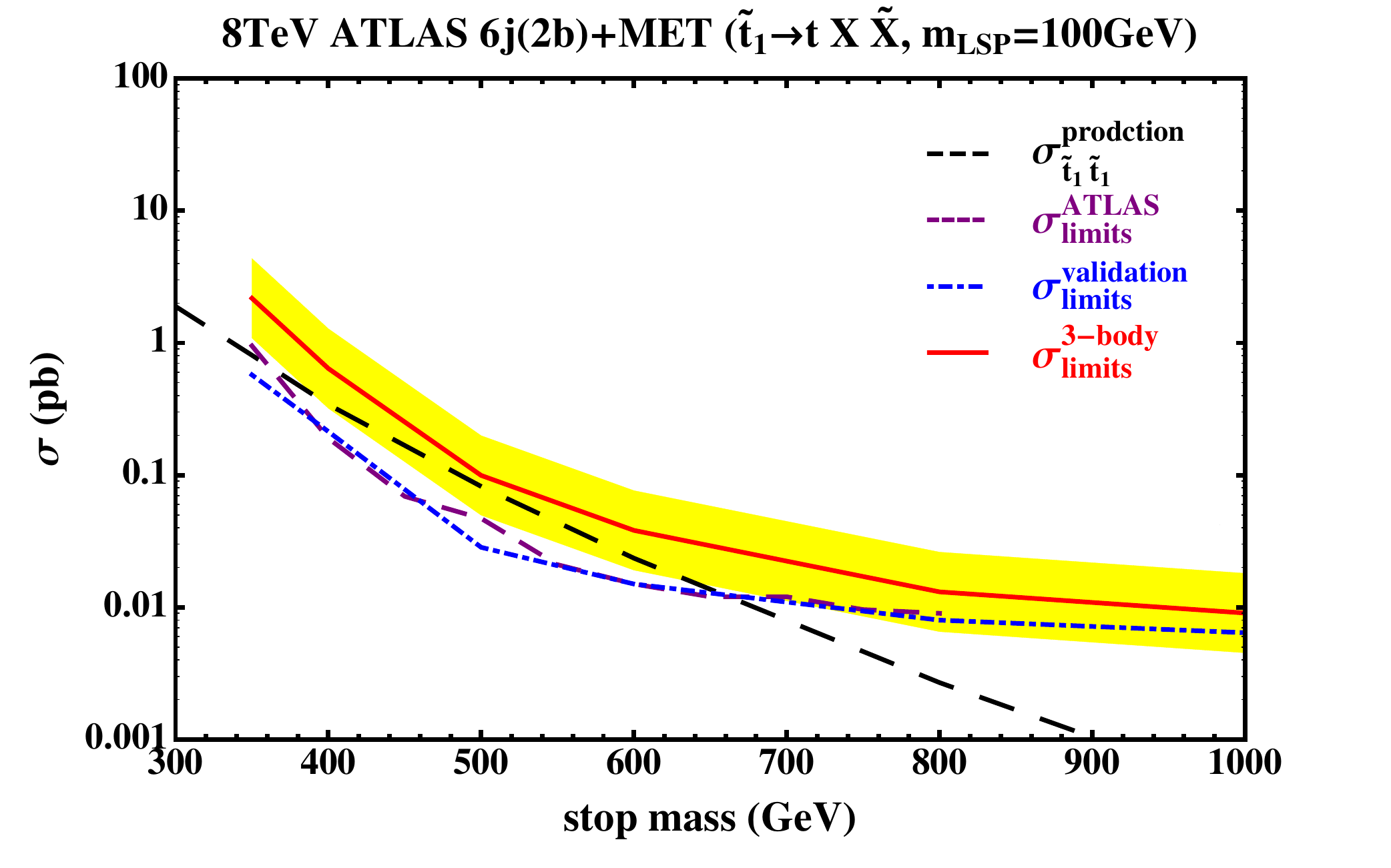} \hskip 0.05\textwidth
\includegraphics[width=0.47\textwidth]{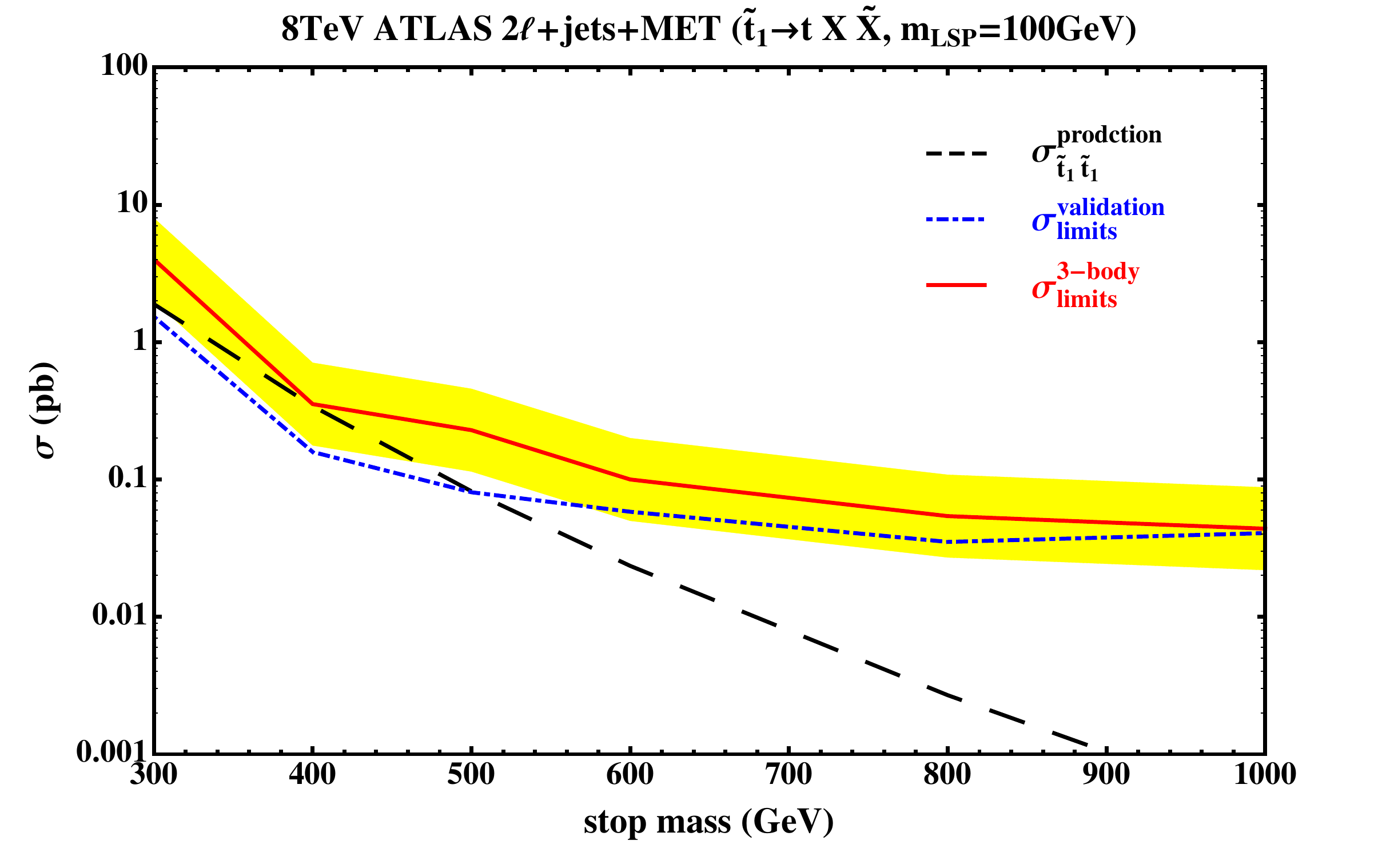} 
\caption{Additional constraints on stops from ATLAS seaches, which are less sensitive than \cite{Chatrchyan:2013xna} presented in Sec.~\ref{sec:limits}. ({\it Left}) ATLAS stop search in the fully hadronic final state  \cite{ATLAS-CONF-2013-024}. ({\it Right}) ATLAS stop search in the dileptonic final state \cite{ATLAS-CONF-2013-048}, which was originally interpreted in the $\tilde t \rightarrow b {\tilde \chi}^+$ topology, and for which reason we do not display our validation limits.}
\end{figure*}

\end{appendix}

\bibliography{missingmet.bib}

\end{document}